\documentclass[journal]{IEEEtran}

\usepackage{url}
\usepackage{cite}
\usepackage{xcolor}
\usepackage{cancel}
\usepackage{siunitx}
\usepackage{fixltx2e}
\usepackage{multirow}
\usepackage{graphicx}
\usepackage{textcomp}
\usepackage{multirow}
\usepackage{stfloats}
\usepackage{algorithm}
\usepackage[T1]{fontenc}
\usepackage{algpseudocode}
\usepackage[10pt]{moresize}
\usepackage[utf8]{inputenc}
\usepackage{mathtools, cuted}
\usepackage{amsmath,amssymb,amsfonts}

\usepackage{array}
\newcommand{\PreserveBackslash}[1]{\let\temp=\\#1\let\\=\temp}
\newcolumntype{C}[1]{>{\PreserveBackslash\centering}p{#1}}
\newcolumntype{R}[1]{>{\PreserveBackslash\raggedleft}p{#1}}
\newcolumntype{L}[1]{>{\PreserveBackslash\raggedright}p{#1}}

\graphicspath{{./figs/}}

\begin{document}

\title{Efficient Hardware Realizations of Feedforward Artificial Neural Networks}

\author{Mohammadreza Esmali Nojehdeh, Sajjad Parvin and Mustafa Altun
	\thanks{This work is an extension of a work originally presented
in ISCAS2020~\cite{aksoy_iscas20}.

The authors are with the Emerging Circuits and Computation Group, Department of Electronics and Communication Engineering, Istanbul Technical University, Maslak, 34469, Istanbul, Turkey (e-mail: \{ nojehdeh, altunmus\}@itu.edu.tr, sajadparvin.stu93@gmail.com).}
}

\maketitle

\begin{abstract}
	This article presents design techniques proposed for efficient hardware implementation of feedforward artificial neural networks (ANNs) under parallel and time-multiplexed architectures. To reduce their design complexity, after the weights of ANN are determined in a training phase, we introduce a technique to find the minimum quantization value used to convert the floating-point weight values to integers. For each design architecture, we also propose an algorithm that tunes the integer weights to reduce the hardware complexity avoiding a loss in the hardware accuracy. Furthermore, the multiplications of constant weights by input variables are implemented under the shift-adds architecture using the fewest number of addition/subtraction operations found by prominent previously proposed algorithms. Finally, we introduce a computer-aided design (CAD) tool, called {\sc simurg}, that can describe an ANN design in hardware automatically based on the ANN structure and the solutions of proposed design techniques and algorithms. Experimental results indicate that the tuning techniques can significantly reduce the ANN hardware complexity under a design architecture and the multiplierless design of ANN can lead to a significant reduction in area and energy consumption, increasing the latency slightly.
\end{abstract}



\section{Introduction}
\label{Sec.intro}

\begin{figure}[t]
	\centerline{\includegraphics[width=9.0cm]{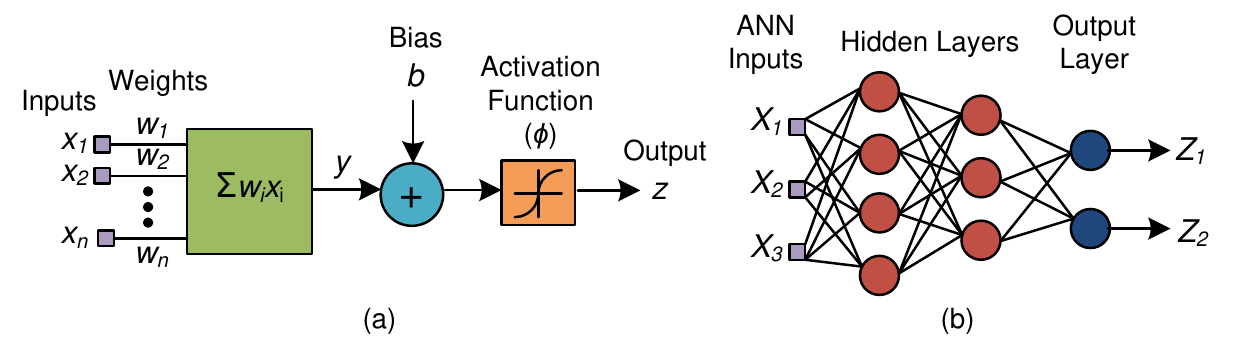}}
	\vspace*{-5mm}
	\caption{(a)~Artificial neuron; (b)~ANN with two hidden layers.}
	\label{fig:neuron}
	\vspace*{-6mm}
\end{figure}

Recent years have seen a tremendous interest in artifical neural networks (ANNs), their successful applications in a wide range of problems, including image recognition~\cite{medical_image} and face detection~\cite{face_detection}, their promising development on graphical processing units (GPUs)~\cite{alexnet}, and their efficient hardware implementations on different design platforms, such as analog, digital, hybrid very large scale integrated circuits (VLSI), and field programmable gate-arrays (FPGAs)~\cite{misra10}. An ANN is a computing system built up by a number of simple and highly interconnected processing elements~\cite{hecht-nielsen90}. As shown in Fig.~\ref{fig:neuron}(a), its fundamental unit, called neuron,  sums the multiplication of weights by input variables, adds the bias value to this summation, and propagates this result to the activation function. While the bias value has the effect of increasing or decreasing the input of the activation function, the activation function limits the amplitude of the neuron output~\cite{haykin99}. Mathematically, the neuron behavior can be defined as \(y = \sum_{i=1}^{n}w_ix_i\) and \(z = \phi(y+b)\) where \(n\) denotes the number of input variables and weights. On the other hand, Fig.~\ref{fig:neuron}(b) presents an ANN design including hidden and output layers where each circle denotes a neuron.

Observe from Fig.~\ref{fig:neuron} that the hardware complexity of an ANN depends heavily on weight and bias values and is dominated by a large number of multiplications of constant weights by input variables. Over the years, many algorithms and design architectures have been introduced to reduce the hardware complexity of ANNs~\cite{courbariaux15,courbariaux16,ding18,hashemi17,hybrid-ann,sarwar16,szabo00,aksoy_iscas20}. In this article, we explore the hardware complexity of ANNs under the parallel and time-multiplexed architectures. Note that a time-multiplexed design, where computations are realized at a time, re-using the computing resources, is preferred to a parallel design in applications with a strict area requirement. However, since the time-multiplexed design needs multiple clock cycles to obtain the final result, it has a higher latency and energy consumption with respect to the parallel design~\cite{parhi99}. To further explore the area versus latency and energy consumption trade-off, in this article, we consider two time-multiplexed architectures. Moreover, since the floating-point multiplication and addition operations occupy larger area and consume more energy than their integer counterparts~\cite{horowitz14}, the floating-point weight and bias values found during training are converted to integers. Since the sizes of integer weight and bias values have a direct impact on the hardware complexity, we introduce a technique that can find the minimum quantization value, sacrificing a little loss in the hardware accuracy. Also, for each design architecture, we propose an algorithm that can tune the weight and bias values such that the hardware complexity is reduced avoiding a loss in the hardware accuracy. Furthermore, since the ANN design includes a large number of multiplications of constant weights by input variables and these weights are determined beforehand, these constant multiplications are realized under the shift-adds architecture using the fewest number of addition/subtraction operations found by previously proposed optimization algorithms~\cite{aksoy10,aksoy12,aksoy_iscas14}. Experimental results clearly indicate that the hardware-aware post-training and the multiplierless design lead to a significant reduction in the ANN hardware complexity with a little loss in the hardware accuracy. Moreover, different design architectures present alternative ANN realizations with different hardware complexity so that a designer can choose the one that fits best in an application.

The rest of this article is organized as follows. Section~\ref{sec:background} presents the background concepts and related work. The parallel and time-multiplexed design architectures are described in Section~\ref{sec:architectures}. The hardware-aware post-training techniques are introduced in Section~\ref{sec:posttraining}. Section~\ref{sec:multiplierless} describes the multiplierless design of ANN. The CAD tool is introduced in Section~\ref{sec:cadtool} and the experimental results are presented in Section~\ref{sec:results}. Finally, Section~\ref{sec:conclusions} concludes the article.

\section{Background}
\label{sec:background}

\subsection{ANN Basics}

Although the design techniques presented in this article can be applied to different ANN architectures, such as convolutional and recurrent, we consider the feedforward ANNs which do not include any feedback loop. Given the ANN structure including the number of inputs, outputs, layers, and neurons in each layer and the activation functions in each layer, the weight and bias values of ANN are determined in a training phase where the error between the desired and actual values is reduced using an iterative optimization algorithm. \mbox{State-of-art} training algorithms~\cite{pytorch,matlab-ann,aksoy_iscas20} consist of efficient techniques on initialization, optimization, and stopping criteria and include a number of activation functions. The training process is generally carried out offline on processors and/or GPUs. In the testing process, the ANN response on the applied inputs is computed using the weight and bias values determined in the training phase. The ANN computation is generally carried out online on a hardware design platform, such as application specific integrated circuits (ASIC) and FPGAs. 

\subsection{Multiplierless Constant Multiplications}
\label{Subsec.MCM}
Multiplication of constants by variable(s) is a ubiquitous and crucial operation in many applications, such as digital signal processing, cryptography, and compilers~\cite{oscar07}. As illustrated in Fig.~\ref{fig:cm_types}, constant multiplications can be categorized in four main classes:

\begin{figure}[t]
	\centerline{\includegraphics[width=9.0cm]{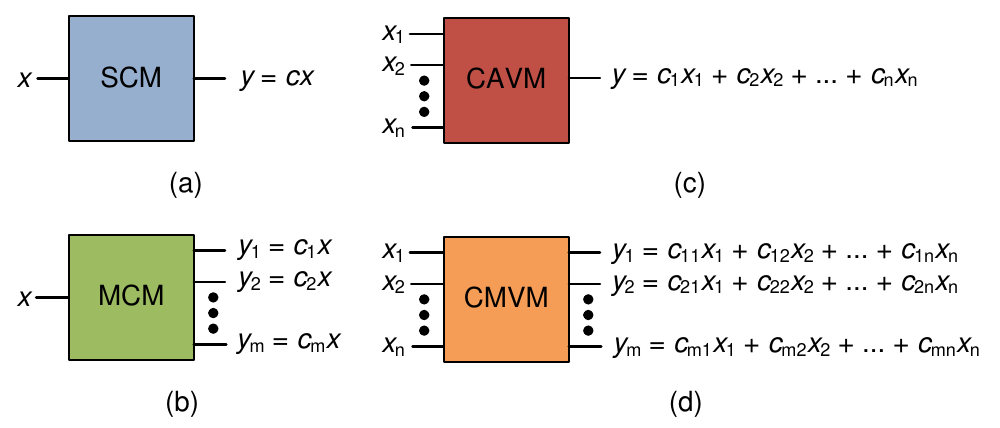}}
	\vspace*{-4mm}
	\caption{Constant multiplications: (a)~single constant multiplication (SCM); (b)~multiple constant multiplication (MCM); (c)~constant array vector multiplication (CAVM); (d)~consant matrix vector multiplication (CMVM).}
	\label{fig:cm_types}
	\vspace*{-6mm}
\end{figure}

\begin{enumerate}
	\item {The \textit{single constant multiplication} (SCM) operation realizes the multiplication of a single constant \(c\) by a single variable \(x\), \textit{i.e.}, \(y=cx\).}
	\item {The \textit{multiple constant multiplication} (MCM) operation computes the multiplication of a set of \(m\) constants \(C\) by a single variable \(x\), \textit{i.e.}, \(y_j=c_jx\) with \(1 \leq j \leq m\).}
	\item {The \textit{constant array-vector multiplication} (CAVM) operation implements the multiplication of a \(1 \times n\) constant array \(C\) by an \(n \times 1\) input vector \(X\), \textit{i.e.}, \(y = \sum_{k}c_kx_k\) with \(1 \leq k \leq n\).}
	\item {The \textit{constant matrix-vector multiplication} (CMVM) operation realizes the multiplication of an \(m \times n\) constant matrix \(C\) by an \(n \times 1\) input vector \(X\), \textit{i.e.}, \(y_j = \sum_{k}c_{jk}x_k\) with \(1 \leq j \leq m\) and \(1 \leq k \leq n\).}
\end{enumerate}

Observe that the CMVM operation is the most general case and corresponds to an SCM operation when both \(m\) and \(n\) are 1, to an MCM operation when \(m > 1\) and \(n\) is 1, and to a CAVM operation when \(m\) is 1 and \(n > 1\).

\begin{figure*}[t]
	\centerline{\includegraphics[width=13.5cm]{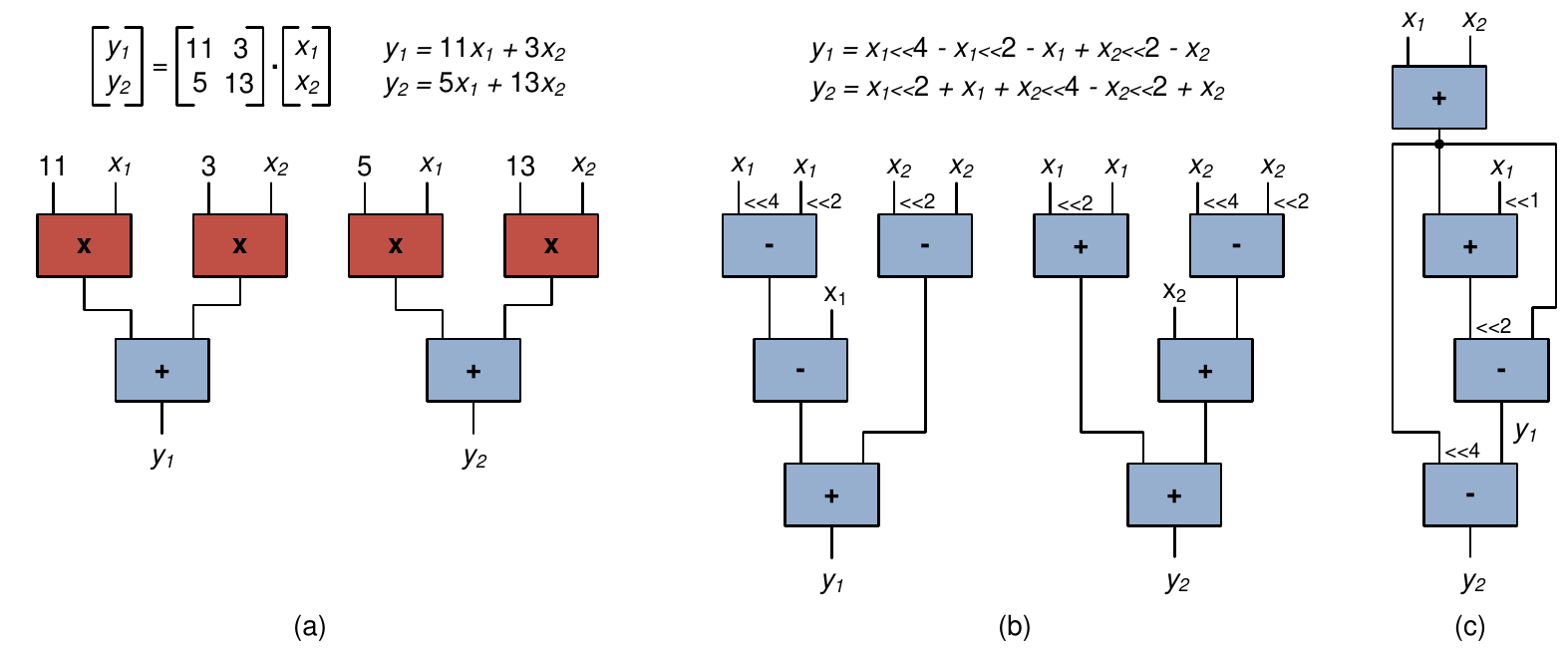}}
	\vspace*{-5mm}
	\caption{(a)~Implementation of a CMVM operation realizing \(y_1=11x_1 + 3x_2\) and \(y_2 = 5x_1 + 13x_2\); Shift-adds implementations: (b)~DBR method~\cite{ercegovac}; (c)~the algorithm of~\cite{aksoy12} optimizing the number of operations.}
	\label{fig:cmvm}
	\vspace*{-6mm}
\end{figure*}

Since the constants are determined beforehand, these constant multiplications can be realized using addition, subtraction, and shift operations under the shift-adds architecture. Note that parallel shifts can be implemented using only wires in hardware without representing any area cost. A straight-forward shift-adds design technique, called the digit-based recoding (DBR)~\cite{ercegovac}, can realize constant multiplications in two steps given as follows: i)~define the constants under a particular number representation, such as binary or canonical signed digit (CSD)\footnote{An integer can be written in CSD using \(n\) digits as \(\sum_{i=0}^{n-1} d_i2^i\), where \(d_i \in \{-1,0,1\}\). The nonzero digits are not adjacent and a constant is represented with a minimum number of nonzero digits under CSD.}; ii)~for the nonzero digits in the representation of constants, shift the input variables according to digit positions and add/subtract the shifted variables with respect to digit values. As a simple example, consider the CMVM operation in Fig.~\ref{fig:cmvm}(a). Its direct realization needs 4 multiplication and 2 addition operations. The DBR method finds a solution with a total number of 8 adders and subtractors when constants are defined under the CSD representation as shown in Fig.~\ref{fig:cmvm}(b). 

The number of adders/subtractors can be further reduced by maximizing the sharing of common partial products among constant multiplications~\cite{boullis05, gustafsson07, voronenko07, aksoy10, aksoy12, aksoy_iscas14}. Returning to our example, the algorithm of~\cite{aksoy12} finds a solution with 4 operations sharing the subexpression \(x_1 + x_2\) as shown in Fig.~\ref{fig:cmvm}(c). Moreover, prominent algorithms, that can find multiplierless designs of constant multiplications taking into account the gate-level area, delay, power dissipation, and throughput of the design, are introduced in~\cite{kang01,demirsoy02,aksoy_dsd10,kumm17}. Furthermore, efficient algorithms are proposed for the multiplierless realization of time-multiplexed constant multiplications in~\cite{demirsoy07, aksoy14, moller16}. 

\subsection{Related Work}

For the multiplierless realization of neural networks, binary neural networks (BNNs), where weights values and activation functions are constrained to be either 1 or -1, were introduced in~\cite{courbariaux16}. It is shown that BNNs drastically reduce the memory size and the number of accesses to the memory during training, and replace multipliers with {\sc xor} operators in hardware. However, they lead to a worse accuracy when compared to conventional neural networks~\cite{ding18}. In~\cite{sarwar16, ding18}, weights of ANNs are determined to include a small number of nonzero digits in training and hence, their multiplications by input variables can be realized using a small number of adders and subtractors. In~\cite{hashemi17}, floating-point weights in each layer are quantized dynamically, fixed-point weights are expressed in binary representation, and the ANN is implemented in a hardware accelerator. The multiplierless hardware realization of ANNs is considered in~\cite{szabo00} where the multiplication of weights by input variables is realized in a bit-serial fashion, defining weights under the CSD representation. In~\cite{aksoy_iscas20}, for the time-multiplexed realization of ANN design, a post-training algorithm, that tunes  weights to reduce the hardware complexity, is introduced and the multiplication of constant weights by input variables in each neuron at each layer is realized under the shift-adds architecture.

Under the time-multiplexed design architecture, the multiply-accumulate (MAC) block is a central operation. To reduce its high latency, a delay-efficient MAC structure, which uses accumulators and carry-save adders, was introduced in~\cite{modified-mac}. Efficient implementation of ANN designs using MAC blocks on FPGAs was introduced in~\cite{fpga-mac}. Recently, MAC blocks have been used in the realization of neuromorphic cores using two models, namely axonal-based and dendritic-based~\cite{hybrid-ann}.

\section{Design Architectures}
\label{sec:architectures}

In this section, we present parallel and time-multiplexed design architectures used to realize ANNs in hardware.

\subsection{Parallel Design}
\label{subsec:parallel}

\begin{figure}[t]
	\vspace*{-2mm}
	\centerline{\includegraphics[width=6.5cm]{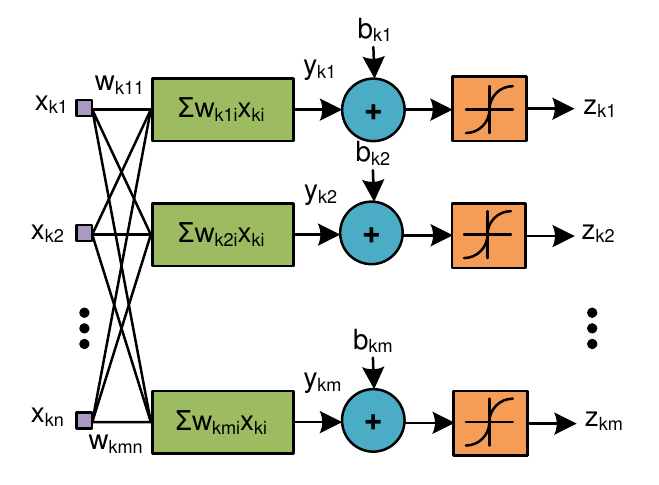}}
	\vspace*{-2mm}
	\caption{Neuron computations at the \(k^{th}\) layer of ANN.}
	\label{fig:ann_parallel}
	\vspace*{-6mm}
\end{figure}

Fig.~\ref{fig:ann_parallel} presents the realization of neuron computations at the \(k^{th}\) layer where \(m\) and \(n\) are the number of outputs (or neurons) and inputs at this layer, respectively. Under the parallel architecture, after the ANN inputs are applied, neuron computations at each layer are obtained concurrently, reaching to the ANN outputs.

\subsection{Time-Multiplexed Design}
\label{subsec:time-multiplexed}

\begin{figure}[t]
	\centerline{\includegraphics[width=8.5cm]{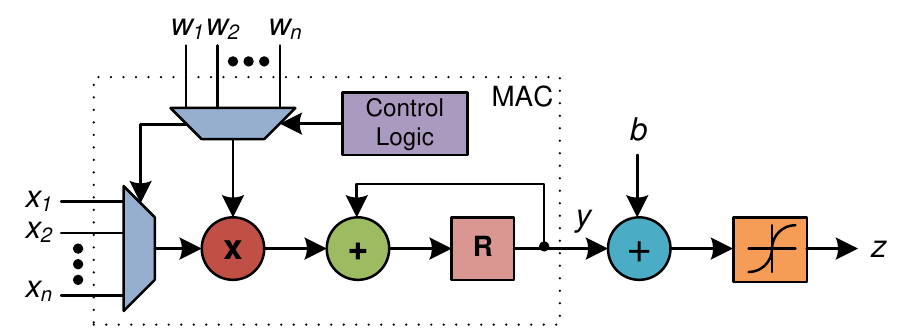}}
	\vspace*{-2mm}
	\caption{Multiply-accumulate (MAC) block in the neuron computation.}
	\label{fig:neuron_mac}
	\vspace*{-6mm}
\end{figure}

The MAC block is a fundamental operation in an ANN design under the time-multiplexed architecture. As shown in Fig.~\ref{fig:neuron_mac}, it can be used to realize the neuron computation given in Fig.~\ref{fig:neuron}(a), re-using the multiplication and addition operations. Observe that the multiplication of a weight by an input variable is realized at a time synchronized by the control block, which is actually a counter, and is added to the accumulated value stored in the register \(R\). In this figure, clock and reset signals are omitted for the sake of clarity. Under this architecture, the neuron computation is obtained after \(n+1\) clock cycles. The design complexity of the MAC block depends on the size of the counter and multiplexers, determined by the number of weights and input variables, on the size of the multiplier, determined by the maximum bitwidths of the input variables and weights, and on the size of adder and register, determined by the bitwidth of the inner product of inputs and weights, \textit{i.e.}, \(y = \sum_{i=1}^{n}w_{i}x_{i}\). 

In this subsection, we present two time-multiplexed architectures to design the whole ANN using MAC blocks. Under the first architecture, called {\sc {\sc smac\_neuron}}, each neuron at each layer is realized using a single MAC block and under the second architecture, called {\sc smac\_ann}, the whole ANN is realized using a single MAC block. In following, these architectures are described in detail.

\subsubsection{{\sc smac\_neuron} Architecture}
\label{subsubsec:smac_neuron}

\begin{figure}[t]
	\centerline{\includegraphics[width=8.5cm]{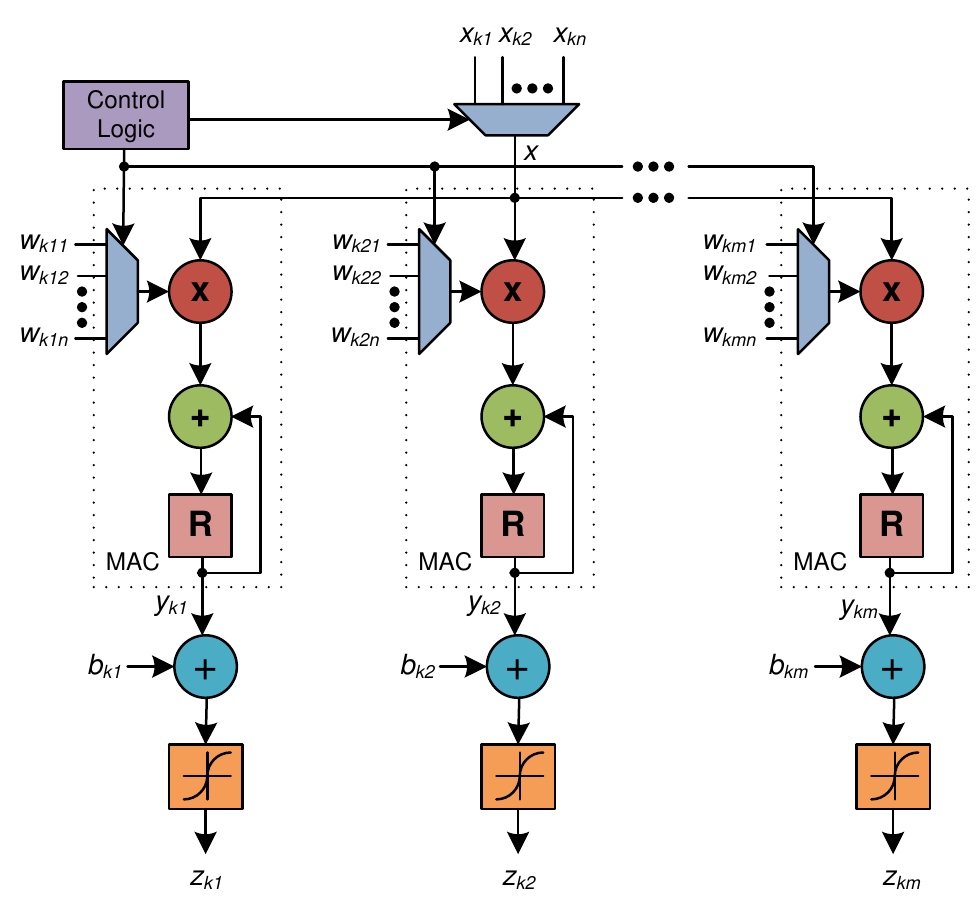}}
	\vspace*{-5mm}
	\caption{Neuron computations at the \(k^{th}\) layer of ANN using MAC blocks.}
	\label{fig:smac_neuron}
	\vspace*{-6mm}
\end{figure}

Fig.~\ref{fig:smac_neuron} presents the neuron computations at the \(k^{th}\) layer of an ANN using \(m\) MAC blocks and a common control block. The control block synchronizes the multiplication of associated weights by input variables. Assuming that an ANN includes \(\eta_i\) neurons at each layer, where \(1 \leq i \leq \lambda\) and \(\lambda\) denotes the number of layers, the required number of MAC blocks is \(\sum_{i}\eta_i\), \textit{i.e.}, the total number of neurons. Note that the complexity of operations and registers in the MAC blocks is determined by the number of inputs and outputs at each layer and the weight values related to each neuron of each layer. The complexity of the control block is determined by the number of inputs at each layer. Since the neuron computations are obtained layer by layer, the neuron computations in a latter layer are started after the neuron computations in a former layer are finished. This is simply done by generating an output signal at each layer indicating that all neuron computations are obtained, which also disables the hardware to do unnecessary computations and enables us to reduce the power dissipation. The computation of the whole ANN with \(\lambda\) layers and \(\iota_i\) inputs at each layer, where \(1 \leq i \leq \lambda\), is obtained after \(\sum_{i}(\iota_i+1)\) clock cycles.

\subsubsection{{\sc smac\_ann} Architecture}
\label{subsubsec:smac_ann}

\begin{figure}[t]
	\centerline{\includegraphics[width=9.0cm]{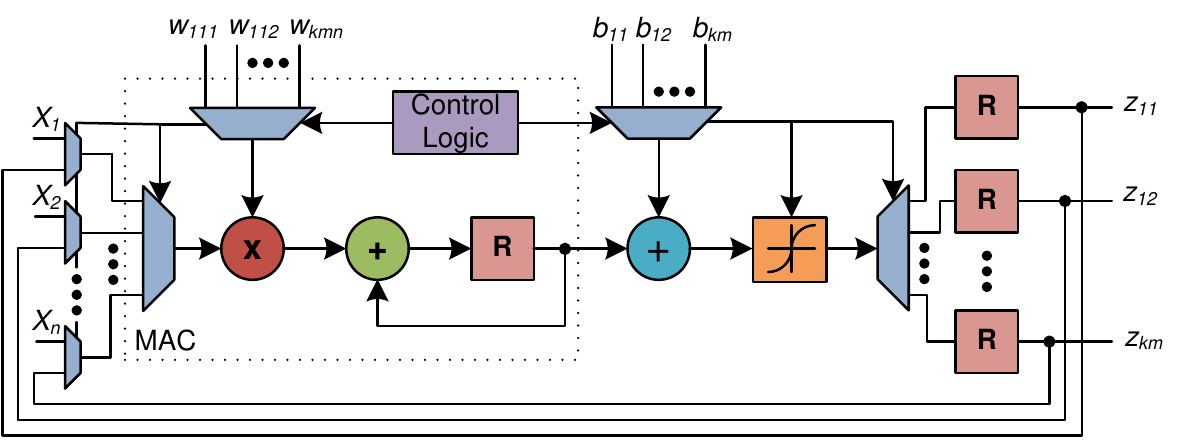}}
	\vspace*{-2mm}
	\caption{ANN design using a single MAC block.}
	\label{fig:smac_ann}
	\vspace*{-6mm}
\end{figure}

Fig.~\ref{fig:smac_ann} shows the ANN design using a single MAC block where clock and reset signals are omitted for the sake of clarity. In this figure, the control block includes three counters to synchronize the multiplication of a weight by an input variable, the addition of a bias value to each inner product, and the application of the activation function. These counters are associated with the number of layers, number of inputs at each layer, and number of outputs (or neurons) at each layer. Note that the variables \(X_1, X_2, \ldots X_n\) denote the primary inputs of ANN and these variables are multiplied by the related weights during the computations at the first layer. While the size of multiplexers for the input variables is determined by the maximum number of inputs at all layers, the size of multiplexers for the weight and bias values are defined by the total number of weight and bias values, respectively. In the MAC block, the size of multiplier is determined by the maximum bitwidth of all input variables and weights and the sizes of adder and register are defined by the maximum bitwidth of the multiplication of weights by input variables in the whole ANN. Moreover, the number of registers used to store the outputs at each layer is determined by the maximum number of outputs at each layer. We note that the computation of the whole ANN with \(\lambda\) layers, \(\iota_i\) inputs at each layer, and \(\eta_i\) neurons at each layer, where \(1 \leq i \leq \lambda\), is obtained after \(\sum_{i}(\iota_i+2)\eta_i\) clock cycles.

\section{Hardware-Aware Post-Training}
\label{sec:posttraining}

In this section, we present a technique proposed for finding the minimum quantization value to convert the floating-point weight and bias values to integers and methods introduced for tuning weight and bias values to reduce the ANN design complexity under the parallel and time-multiplexed architectures.

\subsection{Finding the Minimum Quantization Value}

While converting floating-point weight and bias values found during training to integers, we aim to reduce bitwidths of weight and bias values. To do so, initially, we generate a validation data set, which is used to compute the hardware accuracy, by moving 30\% of the training data set to this set randomly. We note that this validation data set is also used while computing the hardware accuracy during the post-training phase described in following subsections. The proposed technique is described as follows:

\begin{enumerate}
	\item {Set the quantization value, \(q\), and the related ANN accuracy in hardware, \(ha(q)\), to 0.}
	\item {Increase \(q\) value by 1.}
	\item {Convert each floating-point weight and bias value to an integer by multiplying it by \(2^q\) and finding the least integer greater than or equal to this multiplication result.}
	\item {Compute \(ha(q)\) value on the validation data set using the integer weight and bias values.}
	\item {If \(ha(q) > 0\) and \(ha(q)-ha(q-1)\) is greater than 0.1\%, go to Step 2.}
	\item {Otherwise, return \(q\) as the minimum quantization value.}
\end{enumerate}

Observe that we sacrifice maximum 0.1\% loss in the ANN accuracy in hardware computed on the validation data set in order to use small size weight and bias values.

\subsection{Post-Training under the Parallel Architecture}

The main idea for the post-training under the parallel architecture comes from the fact that weight values, which have CSD representations with a small number of nonzero digits, lead to constant multiplications with a small hardware complexity as shown in Fig.~\ref{fig:cmvm}. After weight values are converted to integers using the minimum quantization value \(q\), we check if the least significant nonzero digit in the CSD representation of a weight can be removed avoiding a loss in accuracy. The tuning procedure is described as follows:

\begin{enumerate}
	\item{Set \(ha(q)\) computed in finding the minimum quantization value to the best ANN accuracy in hardware, \(bha\).}
	\item{For each weight \(w\) other than 0, find its CSD representation.}
	\begin{enumerate}
    	\item{Find an alternative weight \(w'\) for \(w\) by removing the least significant nonzero digit in the CSD representation of \(w\) and compute the ANN accuracy in hardware, \(ha'\), when \(w\) is replaced by \(w{'}\).}
    	\item{If \(ha' \geq bha\), replace \(w\) by \(w'\) and update \(bha\).}
    \end{enumerate}
	\item{If at least one weight is replaced by an alternative one, go to Step 2.}
	\item {Otherwise, return weight values.}
\end{enumerate}

Note that in Step 2 of the tuning procedure, the number of nonzero digits in the CSD representation of an alternative weight, which replaces the original weight, is always less than that of the original weight.

\subsection{Post-Training under the Time-Multiplexed Architectures}
\label{Subsec.postrain_MCM}
The main idea for the post-training under the time-multiplexed architectures comes from the fact that in the MAC block shown in Fig.~\ref{fig:neuron_mac}, if all weights are multiples of \(2^k\), where \(k > 0\), the inner product can be realized as \(y = (\sum_{i=1}^{n}c_{i}x_{i})\ll k\), where \(c_i = w_i/2^k\) and \(\ll\) stands for the left shift operation. Thus, bitwidths of constant weights to be multiplied by input variables in the MAC block, and consequently, sizes of the multiplier, adder, and register can be reduced. After weight values are converted to integers using the minimum quantization value \(q\), under the {\sc smac\_neuron} architecture, for each neuron, we aim to maximize the smallest left shift value among all its weights, denoted as \(sls\), while avoiding a decrease in the ANN accuracy. As a simple example, the \(sls\) value for the integer values \(20~=~5~\ll~2\), \(24 = 3 \ll 3\), and \(26 = 13 \ll 1\), is computed as 1. The tuning procedure presented for the {\sc smac\_neuron} architecture~\cite{aksoy_iscas20} is described as follows:

\begin{enumerate}
	\item{Set \(ha(q)\) computed in finding the minimum quantization value to the best ANN accuracy in hardware, \(bha\).}
	\item{For each neuron in the ANN design, compute the \(sls\) value for its weights.}
	\begin{enumerate}
		\item{For each weight associated the \(m^{th}\) neuron and \(n^{th}\) input at the \(k^{th}\) level, \(w_{kmn}\), find its largest left shift value, \(lls\)}.
		\item{If \(lls\) is equal to \(sls\), determine the first possible weight as \(pw_1 = w_{kmn} - (w_{kmn} \mod 2^{lls+1})\). If the bitwidth of \(pw_1\) is less than or equal to the maximum  bitwidth of weights associated with \(N_i\), compute the ANN accuracy in hardware, \(ha_1\), when \(w_{kmn}\) is replaced by \(pw_1\).  Similarly, determine the second possible weight as \(pw_2 = pw_1 + 2^{lls+1}\) and compute the related ANN hardware accuracy, \(ha_2\).}
		\item{If \(max(ha_1, ha_2) \geq bha\), replace \(w_{kmn}\) by the possible weight that leads to the maximum ANN accuracy in hardware and update \(bha\) accordingly.}
		\item{Otherwise, assuming that \(w_{kmn}\) is replaced by the possible weight that leads to the maximum ANN accuracy in hardware, change the bias value of the neuron, \(b_{km}\), in between \([b_{km}-4, b_{km}+4]\) and compute the ANN accuracy in hardware. If the ANN accuracy in hardware in one of these cases is greater than or equal to \(bha\), update the values of \(w_{kmn}\), \(b_{km}\), and \(bha\) accordingly.}
	\end{enumerate}
	\item{If \(sls\) value of any neuron is improved, go to Step 2.}
	\item{Otherwise, return weight and bias values.}
\end{enumerate}

Note that in Step 2 of the tuning procedure, a possible weight, which replaces the original weight, has always the largest left shift value greater than that of the original weight.

For the {\sc smac\_ann} architecture, we apply a similar procedure where the increment of the smallest left shift of all ANN weights is aimed in the MAC block as shown in Fig.~\ref{fig:smac_ann}. 

\section{ANNs Under the Shift-Adds Architecture}
\label{sec:multiplierless}

This section presents the multiplierless realizations of ANN designs under the parallel and time-multiplexed architectures.

\subsection{Multiplierless ANN Design under the Parallel Architecture}

A straight-forward way for the multiplierless realization of ANN under the parallel architecture is to describe each inner product at each layer, \textit{i.e.}, \(y_{k1}, y_{k2}, \ldots y_{kn}\) shown in Fig.~\ref{fig:ann_parallel}, as a CAVM operation and to implement each CAVM block independently under the shift-adds architecture. We use the algorithm of~\cite{aksoy_iscas14} to optimize the number of adders/subtractors in the multiplierless designs of these CAVM blocks.

\begin{figure}[t]
	\centerline{\includegraphics[width=8.5cm]{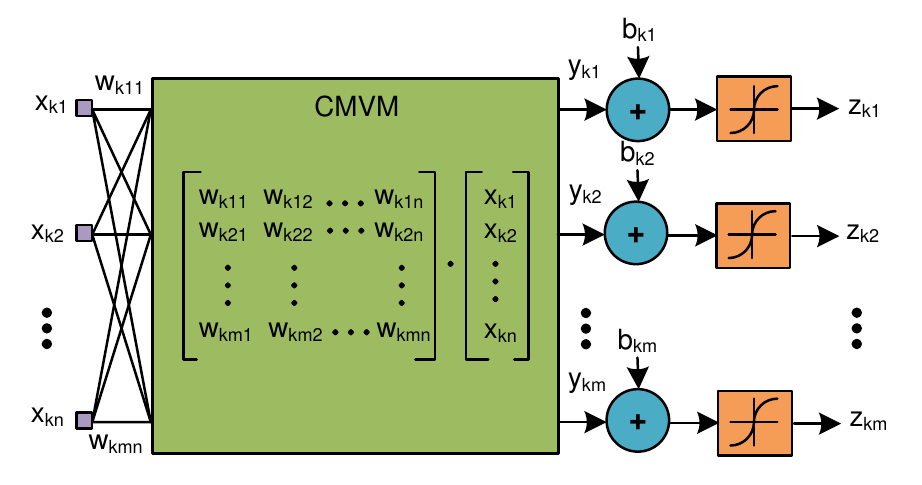}}
	\vspace*{-5mm}
	\caption{Neuron computations at the \(k^{th}\) layer using a CMVM block.}
	\label{fig:ann_parallel_cmvm}
	\vspace*{-6mm}
\end{figure}

As shown in Fig.~\ref{fig:ann_parallel_cmvm}, all inner products at the \(k^{th}\) layer can be described as a CMVM operation and the number of adders/subtractors in the multiplierless realization of the CMVM block can be reduced using the algorithm of~\cite{aksoy12}. Thus, the possible sharing of subexpressions can be increased, reducing the number of adders and subtractors in the multiplierless ANN design.

\subsection{Multiplierless ANN Design under the Time-Multiplexed Architectures}

\begin{figure}[t]
	\centerline{\includegraphics[width=8.5cm]{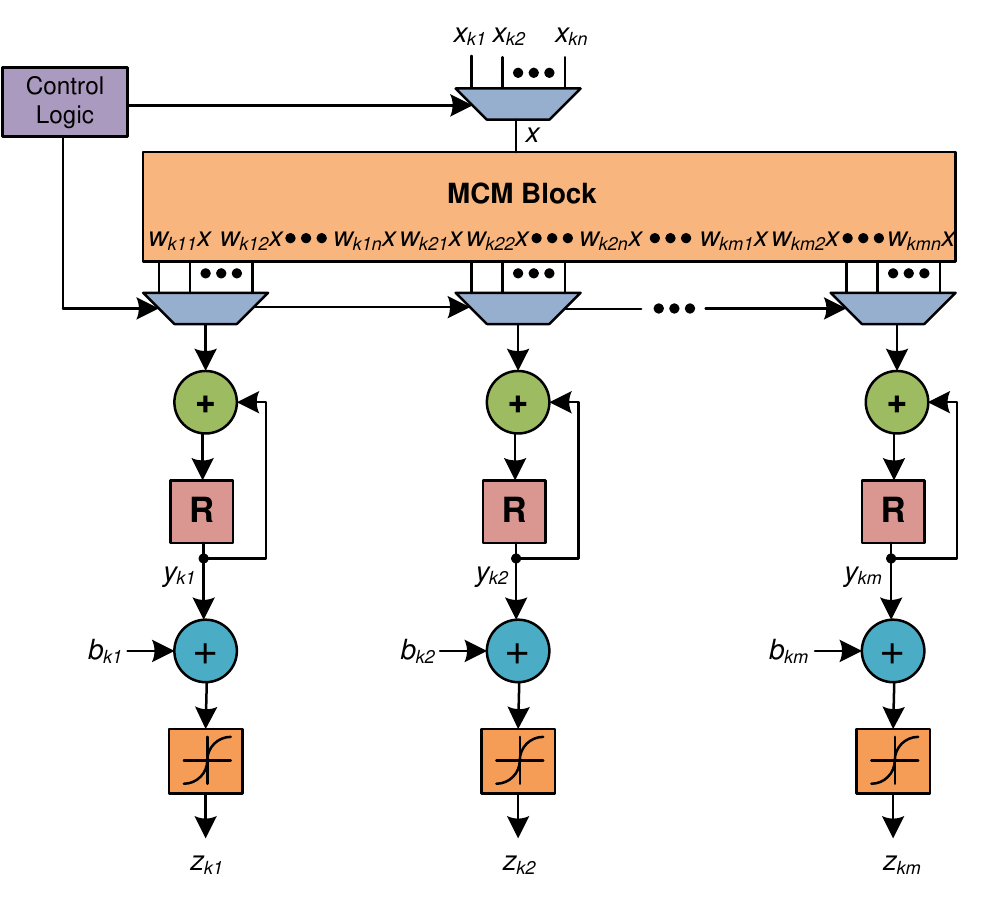}}
	\vspace*{-5mm}
	\caption{Multiplierless realization of neuron computations at the \(k^{th}\) layer under the {\sc smac\_neuron} architecture.}
	\label{fig:smac_neuron_mulless}
	\vspace*{-6mm}
\end{figure}

Under the {\sc smac\_neuron} architecture, the multiplications of associated weights by input variables at the \(k^{th}\) layer shown in Fig.~\ref{fig:smac_neuron} can be computed in an MCM block and can be diverted to the corresponding adders using multiplexers as shown in Fig.~\ref{fig:smac_neuron_mulless}. Rather than using an MCM block for each neuron, a single MCM block, which realizes the multiplication of all weights in a layer by an input variable, is used to increase the sharing of partial products, and thus, to reduce the required number of adders/subtractors. The exact algorithm of~\cite{aksoy10} is used to find the shift-adds realization of the MCM block using a minimum number of adders/subtractors.

Similarly, the multiplierless realization of ANN under the {\sc smac\_ann} architecture presented in Fig.~\ref{fig:smac_ann} can be obtained when the multiplication of all weight values by the selected input variable is implemented using an MCM block. However, since one multiplier is replaced by a large number of adders/subtractors, such a multiplierless realization increases the hardware complexity significantly.

\section{SIMURG: The CAD Tool}
\label{sec:cadtool}

In this section, we present our CAD tool called {\sc simurg} developed to generate automatically the hardware description of an ANN under the design architectures given in Section~\ref{sec:architectures} using the post-training methods mentioned in Section~\ref{sec:posttraining} and the multiplierless design techniques described in Section~\ref{sec:multiplierless}. 

Given the ANN structure, including the number of inputs, outputs, hidden layers, and neurons in the hidden layers and the type of activation function of neurons in each layer, initially, the ANN is trained using a state-of-art technique and the weight and bias values are determined. In this study, we used three techniques to train the ANN, namely, our training algorithm~\cite{aksoy_iscas20}, called {\sc zaal}, Pytorch~\cite{pytorch}, and {\sc matlab} neural network toolbox~\cite{matlab-ann}. {\sc zaal} includes the conventional and stochastic gradient descent methods, and the Adam optimizer~\cite{adam14} as an iterative optimization algorithm. It has different weight initialization techniques, such as Xavier~\cite{xavier10}, He~\cite{he15}, and a fully random method. It includes several stopping criteria, \textit{e.g.}, the number of iterations, early stopping using validation data set, and saturation of loss function. It can define a number of activation functions for neurons in each layer, namely sigmoid, hard sigmoid (hsig), hyperbolic tangent, hard hyperbolic tangent (htanh), linear (lin), rectified linear unit (ReLU), saturating linear (satlin), and softmax~\cite{nwankpa18}.

After floating-point weight and bias values determined during the training phase are converted to integers with a given quantization value and/or the design techniques introduced in Section~\ref{sec:posttraining} are applied, the ANN design is described in hardware automatically based on the ANN structure given to a training algorithm, the integer weight and bias values, and the design architecture, \textit{i.e.}, parallel, {\sc smac\_neuron}, or {\sc smac\_ann}. The activation functions used in {\sc simurg} are \textit{hsig}, \textit{htanh}, \textit{lin}, \textit{ReLU}, and \textit{satlin} due to their simplicity in hardware. The tool can define the multiplication of constant weights by input variables in a behavioral fashion. Also, it can find the multiplierless realizations of these constant multiplications as described in Section~\ref{sec:multiplierless}. The tool also generates a test-bench and necessary files to verify the ANN design and the synthesis scripts automatically. The {\sc simurg} tool with its limited number of functions is available at \mbox{https://github.com/leventaksoy/ANNs}.

\section{Experimental Results}
\label{sec:results}

In this work, we used the pen-based handwritten digit recognition problem~\cite{alimoglu97} as an application. In the convolutional neural network design of this application, we implemented 5 feedforward ANN structures with different number of layers and number of neurons in layers, denoted as \(p_{in} \textrm{--} \eta_{1} \textrm{--} \eta_{2} \textrm{--} \ldots \textrm{--} \eta_{\lambda}\), where \(p_{in}\) stands for the number of ANN primary inputs, which is equal to 16, and \(\eta_{k}\), where \(1 \leq k \leq \lambda\), indicates the number of neurons in the \(k^{th}\) layer. Note that the activation function of each neuron in the hidden and output layers in training (hardware) was respectively \textit{htanh} (\textit{htanh}) and \textit{sigmoid} (\textit{hsig}) in our training algorithm {\sc zaal} and {\sc pytorch}. In {\sc matlab}, it was respectively as \textit{tanh} (\textit{htanh}) and \textit{satlin} (\textit{satlin}). The activation functions were determined based on the software test accuracy found in training. The ANNs were trained using 7494 data and tested using 3498 data. We note that each training algorithm was run 30 times and weight and bias values, that yielded the best accuracy value, were chosen. Table~\ref{tab:ttools} presents the training and hardware design details on different ANN design structures. In this table, \textit{sta}, \textit{hta}, and \textit{tnzd} denote the software test accuracy, hardware test accuracy, and the total number of nonzero digits in the CSD representations of integer weights and bias values, respectively. Floating-point weight and bias values were converted to integers using the minimum quantization value determined as described in Section~\ref{sec:posttraining}. In the ANN hardware design, bitwidths of ANN inputs and outputs at each layer were determined as 8. 

\begin{table}[t]
	\centering
	\scriptsize
	\caption{Details of ANNs on training and hardware design.}
	\vspace*{-3mm}
	\begin{tabular}{|L{12.3mm}||C{3.5mm}|C{3.5mm}|C{3.5mm}||C{3.5mm}|C{3.5mm}|C{3.5mm}||C{3.5mm}|C{3.5mm}|C{3.5mm}|}
		\hline
		\multirow{2}{*}{Structure} & \multicolumn{3}{c||}{{\sc zaal}} & \multicolumn{3}{c||}{{\sc pytorch}~\cite{pytorch}} & \multicolumn{3}{c|}{{\sc matlab}~\cite{matlab-ann}}\\
		\cline{2-10}
		& sta & hta & tnzd & sta & hta & tnzd & sta & hta & tnzd\\
		\hline \hline
		16-10       & 84.6 & 86.0 & 431  & 85.5 & 85.1 & 374  & 89.1 & 89.3 & 374  \\
		16-10-10    & 94.1 & 93.6 & 855  & 95.9 & 95.2 & 950  & 95.9 & 95.9 & 857  \\
		16-16-10    & 96.0 & 95.9 & 1245 & 95.6 & 95.6 & 1338 & 96.9 & 95.0 & 1291 \\
		16-10-10-10 & 94.7 & 94.0 & 1121 & 95.8 & 95.6 & 1190 & 96.4 & 94.7 & 1121 \\
		16-16-10-10 & 96.6 & 96.6 & 1432 & 96.7 & 96.7 & 1608 & 96.6 & 95.2 & 1560 \\
		\hline \hline
		Average     & 93.2 & 93.2 & 1017 & 93.9 & 93.6 & 1092 & 95.0 & 94.0 & 1041 \\
		\hline
	\end{tabular}
	\label{tab:ttools}
	\vspace*{-6mm}
\end{table}

Observe from Table~\ref{tab:ttools} that different training algorithms lead to ANN designs with different hardware accuracy and integer weights and bias values. However, they yield software test accuracy values close to hardware test accuracy values. Note that the difference between the software and hardware test accuracy is due to the quantization value, bitwidths of ANN inputs and outputs at each layer, and different activation functions used in training and hardware design.

\begin{figure*}[t]
	\centerline{\includegraphics[width=18.5cm]{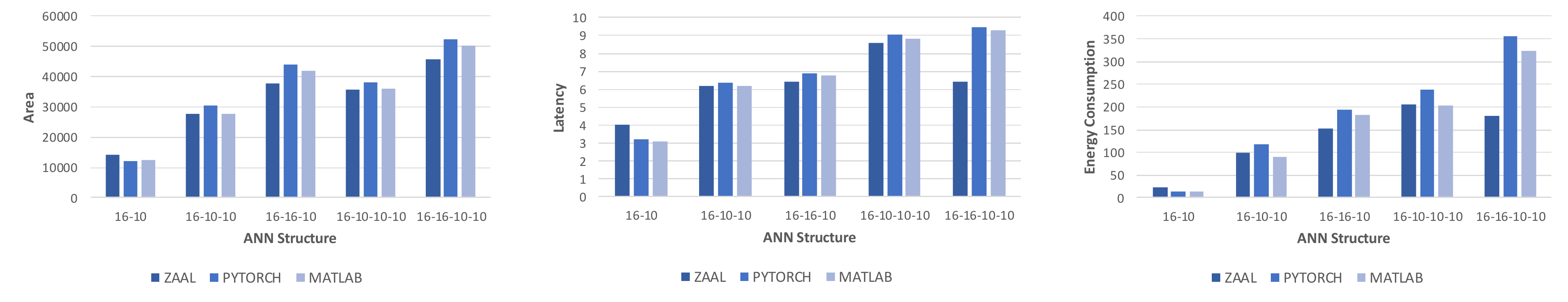}}
	\vspace*{-2mm}
	\caption{ANN designs under the parallel architecture when constant multiplications are described in a behavioral fashion and no post-training is applied.}
	\label{fig:ttools_parallel}
	\vspace*{-4mm}
\end{figure*}

\begin{figure*}[t]
	\centerline{\includegraphics[width=18.5cm]{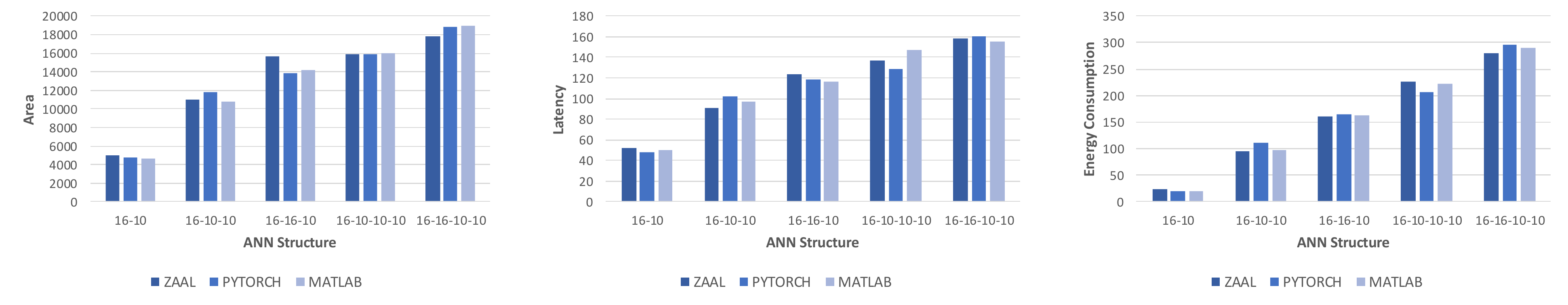}}
	\vspace*{-2mm}
	\caption{ANN designs under the {\sc smac\_neuron} architecture when constant multiplications are described in a behavioral fashion and no post-training is applied.}
	\label{fig:ttools_smac_neuron}
	\vspace*{-4mm}
\end{figure*}

\begin{figure*}[t]
	\centerline{\includegraphics[width=18.5cm]{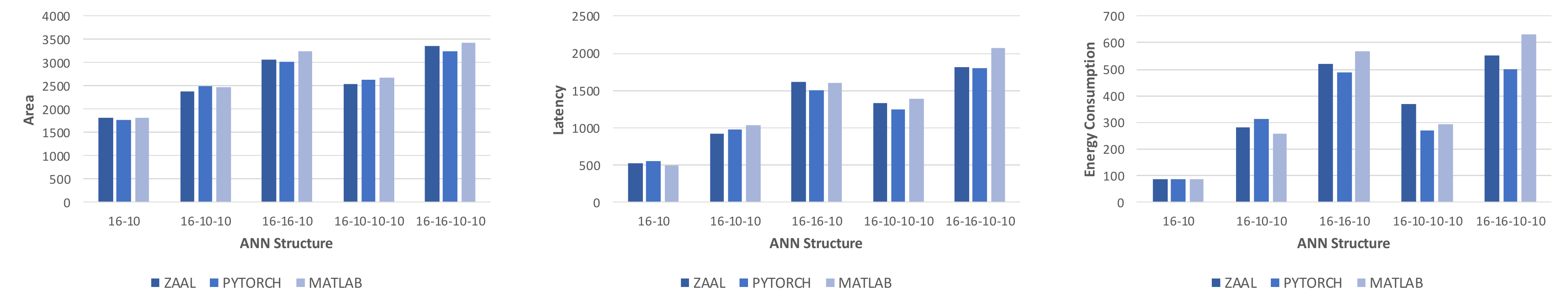}}
	\vspace*{-2mm}
	\caption{ANN designs under the {\sc smac\_ann} architecture when constant multiplications are described in a behavioral fashion and no post-training is applied.}
	\label{fig:ttools_smac_ann}
	\vspace*{-6mm}
\end{figure*}

In this work, we present gate-level results of ANN designs implemented in three different architectures, namely parallel, {\sc smac\_neuron}, and {\sc smac\_ann}, as described in Section~\ref{sec:architectures}. In parallel designs, to make a fair comparison with time-multiplexed designs, flip-flops were added to outputs of the ANN design. ANN designs were described in Verilog and were synthesized using the Cadence RTL Compiler with the TSMC 40nm design library. 

In order to explore the impact of a design architecture on the ANN hardware complexity, Figs.~\ref{fig:ttools_parallel}-\ref{fig:ttools_smac_ann} present respectively area (in \(\mu m^2\)), latency (in \(ns\)), and energy consumption (in \(pJ\)) results of ANN designs under the parallel, {\sc smac\_neuron}, and {\sc smac\_ann} architectures where no post-training technique is applied and constant multiplications are described in a behavioral fashion. Note that the latency is computed as the multiplication of the clock period by the number of clock cycles to obtain the ANN output. The clock period was reduced using the retiming technique in the synthesis tool iteratively. The switching activity data required for the computation of power dissipation was generated using the test data in simulation. This test data set was also used to verify the ANN design. Energy consumption is computed as the multiplication of latency and power dissipation. 

Observe that weight and bias values found by different training algorithms lead to ANN designs with different hardware complexity where their impact is clearly observed on ANN designs under the parallel architecture since there exist a large number of constant multiplications. On the other hand, while ANN designs under the {\sc smac\_ann} architecture have the smallest area, the ones under the parallel architecture occupy the largest area. However, the latency of ANN designs under the parallel architecture is significantly smaller than those of ANN designs under the time-multiplexed architectures. Moreover, ANN designs under the {\sc smac\_ann} architecture consume the most energy. Note that area, latency, and energy consumption values of ANN designs under the {\sc smac\_neuron} architecture are in between those of ANN designs under the parallel and {\sc smac\_ann} architectures.

\begin{table}[t]
	\centering
	\scriptsize
	\caption{Details of ANNs designs under the parallel architecture after the post-training phase.}
	\vspace*{-3mm}
	\begin{tabular}{|L{12.3mm}||C{3.5mm}|C{3.5mm}|C{3.5mm}||C{3.5mm}|C{3.5mm}|C{3.5mm}||C{3.5mm}|C{3.5mm}|C{3.5mm}|}
		\hline
		\multirow{2}{*}{Structure} & \multicolumn{3}{c||}{{\sc zaal}} & \multicolumn{3}{c||}{{\sc pytorch}~\cite{pytorch}} & \multicolumn{3}{c|}{{\sc matlab}~\cite{matlab-ann}}\\
		\cline{2-10}
		& hta & tnzd & CPU & hta & tnzd & CPU & hta & tnzd & CPU \\
		\hline \hline
16-10       & 86.2 & 224 & 111  & 86.0 & 184 & 136  & 89.0 & 264 & 113  \\
16-10-10    & 92.9 & 426 & 338  & 93.9 & 421 & 334  & 95.3 & 416 & 342  \\
16-16-10    & 95.1 & 425 & 851  & 94.7 & 469 & 996  & 94.9 & 609 & 590  \\
16-10-10-10 & 93.4 & 456 & 912  & 95.0 & 498 & 931  & 94.9 & 550 & 488  \\
16-16-10-10 & 95.2 & 544 & 1127 & 94.4 & 615 & 1254 & 95.1 & 693 & 1207 \\
		\hline \hline
Average     & 92.6 & 415 & 668  & 92.8 & 437 & 730  & 93.8 & 506 & 548  \\
		\hline
	\end{tabular}
	\label{tab:tuning_parallel}
	\vspace*{-4mm}
\end{table}

\begin{table}[t]
	\centering
	\scriptsize
	\caption{Details of ANNs designs under the {\sc smac\_neuron} architecture after the post-training phase.}
	\vspace*{-3mm}
	\begin{tabular}{|L{12.3mm}||C{3.5mm}|C{3.5mm}|C{3.5mm}||C{3.5mm}|C{3.5mm}|C{3.5mm}||C{3.5mm}|C{3.5mm}|C{3.5mm}|}
		\hline
		\multirow{2}{*}{Structure} & \multicolumn{3}{c||}{{\sc zaal}} & \multicolumn{3}{c||}{{\sc pytorch}~\cite{pytorch}} & \multicolumn{3}{c|}{{\sc matlab}~\cite{matlab-ann}}\\
		\cline{2-10}
		& hta & tnzd & CPU & hta & tnzd & CPU & hta & tnzd & CPU \\
		\hline \hline
16-10       & 86.6 & 279 & 108 & 84.9 & 272 & 78   & 88.8 & 301 & 87  \\
16-10-10    & 93.5 & 550 & 515 & 94.4 & 563 & 552  & 95.3 & 518 & 651  \\
16-16-10    & 95.9 & 694 & 644 & 95.0 & 753 & 765  & 94.9 & 813 & 670  \\
16-10-10-10 & 93.5 & 755 & 544 & 95.7 & 699 & 1259 & 95.0 & 726 & 813  \\
16-16-10-10 & 95.6 & 816 & 789 & 95.9 & 918 & 1489 & 95.3 & 991 & 981  \\
		\hline \hline
Average     & 93.0 & 618 & 520 & 93.2 & 641 & 829  & 93.9 & 669 & 641  \\
		\hline
	\end{tabular}
	\label{tab:tuning_smac_neuron}
	\vspace*{-7mm}
\end{table}

\begin{table}[t]
	\centering
	\scriptsize
	\caption{Details of ANNs designs under the {\sc smac\_ann} architecture after the post-training phase.}
	\vspace*{-3mm}
	\begin{tabular}{|L{12.3mm}||C{3.5mm}|C{3.5mm}|C{3.5mm}||C{3.5mm}|C{3.5mm}|C{3.5mm}||C{3.5mm}|C{3.5mm}|C{3.5mm}|}
		\hline
		\multirow{2}{*}{Structure} & \multicolumn{3}{c||}{{\sc zaal}} & \multicolumn{3}{c||}{{\sc pytorch}~\cite{pytorch}} & \multicolumn{3}{c|}{{\sc matlab}~\cite{matlab-ann}}\\
		\cline{2-10}
		& hta & tnzd & CPU & hta & tnzd & CPU & hta & tnzd & CPU \\
		\hline \hline
16-10       &  86.1 & 362 & 32  & 85.7 & 318  & 24  & 89.2 & 339  & 37 \\
16-10-10    &  93.5 & 611 & 192 & 94.8 & 615  & 387 & 95.7 & 579  & 170 \\
16-16-10    &  95.9 & 829 & 253 & 95.4 & 781  & 457 & 94.9 & 878  & 388 \\
16-10-10-10 &  93.6 & 770 & 381 & 95.8 & 1057 & 92  & 95.1 & 899  & 168 \\
16-16-10-10 &  96.4 & 960 & 360 & 96.5 & 1426 & 156 & 95.7 & 1041 & 618 \\
		\hline \hline
Average     &  93.1 & 706 & 244 & 93.6 & 839  & 223 & 94.1 & 747  & 276 \\
		\hline
	\end{tabular}
	\label{tab:tuning_smac_ann}
	\vspace*{-7mm}
\end{table}

Tables~\ref{tab:tuning_parallel}-\ref{tab:tuning_smac_ann} present details on the ANN designs under the parallel, {\sc smac\_neuron}, and {\sc smac\_ann} architectures after the post-training phase, respectively. In these tables, \textit{CPU} denotes the run-time of post-training algorithms given in seconds. Observe from Table~\ref{tab:ttools} and these tables that the high-level hardware cost value, \textit{i.e.}, \textit{tnzd}, is reduced significantly with a little loss in the hardware accuracy. However, there are cases where the hardware test accuracy is increased after the post-training phase, such as all \mbox{16-10} ANN designs trained using {\sc zaal} and \mbox{16-10-10-10} ANN designs under {\sc smac\_neuron} and {\sc smac\_ann} architectures trained using {\sc pytorch}. 

\begin{figure*}[t]
	\centerline{\includegraphics[width=18.5cm]{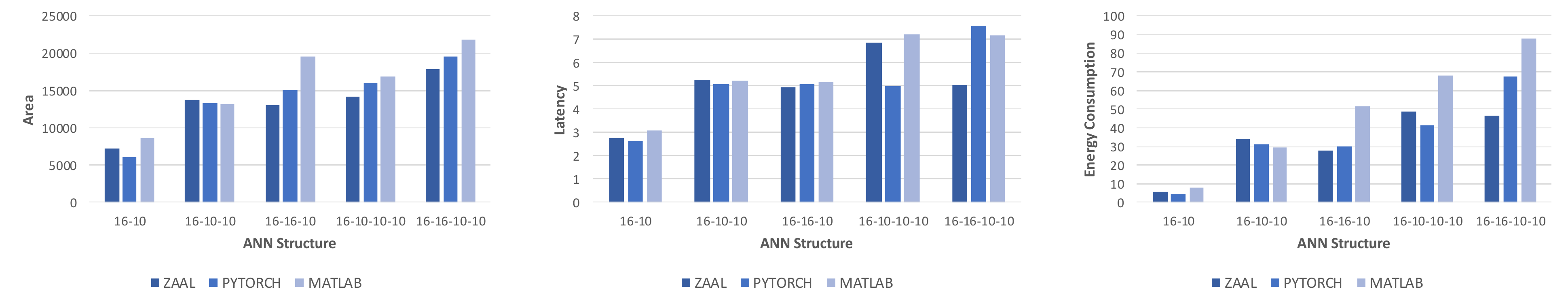}}
	\vspace*{-2mm}
	\caption{ANN designs under the parallel architecture when constant multiplications are described in a behavioral fashion after the post-training phase.}
	\label{fig:tuning_parallel}
	\vspace*{-4mm}
\end{figure*}

\begin{figure*}[t]
	\centerline{\includegraphics[width=18.5cm]{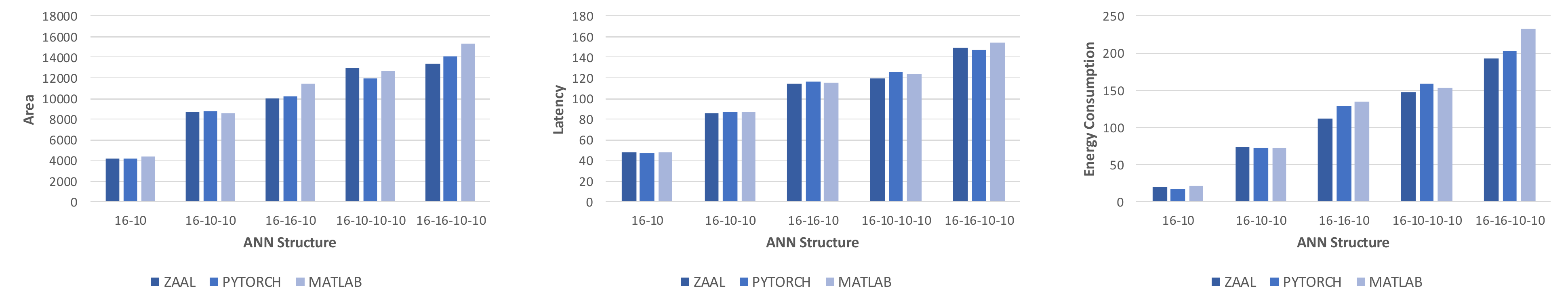}}
	\vspace*{-2mm}
	\caption{ANN designs under the {\sc smac\_neuron} architecture when constant multiplications are described in a behavioral fashion after the post-training phase.}
	\label{fig:tuning_smac_neuron}
	\vspace*{-4mm}
\end{figure*}

\begin{figure*}[t]
	\centerline{\includegraphics[width=18.5cm]{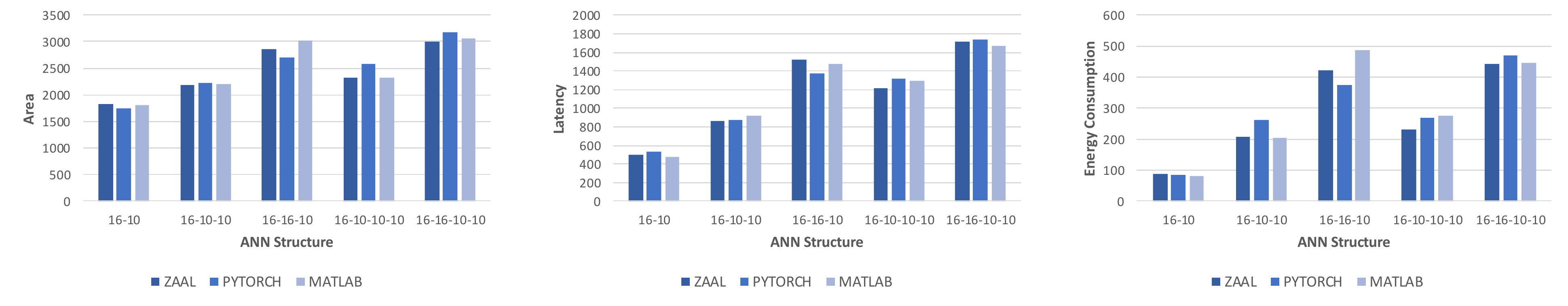}}
	\vspace*{-2mm}
	\caption{ANN designs under the {\sc smac\_ann} architecture when constant multiplications are described in a behavioral fashion after the post-training phase.}
	\label{fig:tuning_smac_ann}
	\vspace*{-6mm}
\end{figure*}

In order to explore the impact of tuning algorithms on the ANN hardware complexity, Figs.~\ref{fig:tuning_parallel}-\ref{fig:tuning_smac_ann} present respectively gate-level results of ANN designs obtained after weight and bias values are tuned to reduce the hardware complexity of ANN designs under the parallel, {\sc smac\_neuron}, and {\sc smac\_ann} architectures where constant multiplications are described in a behavioral fashion.

Observe from Figs.~\ref{fig:ttools_parallel} and~\ref{fig:tuning_parallel} that the use of a tuning algorithm can significantly reduce the hardware complexity of ANN designs under the parallel architecture. When compared to ANN designs realized without the post-training phase, we note that the maximum reduction on area, latency, and energy consumption is found as 65\%, 44\%, and 84\% on the \mbox{16-16-10} ANN design trained using {\sc zaal}, \mbox{16-10-10-10} and \mbox{16-16-10} ANN designs trained using {\sc pytorch}, respectively. Similar results are observed on time-multiplexed designs. On the ANN designs under the {\sc smac\_neuron} architecture, the maximum reduction on area, latency, and energy consumption is found as 35\%, 15\%, and 34\% on the \mbox{16-16-10} ANN design trained using {\sc zaal}, \mbox{16-10-10-10} ANN design trained using {\sc matlab}, and \mbox{16-10-10-10} ANN designs trained using {\sc zaal}, respectively. Finally, on the ANN designs under the {\sc smac\_ann} architecture, the maximum reduction on area, latency, and energy consumption is found as 12\%, 19\%, and 37\% on the \mbox{16-10-10-10} ANN design trained using {\sc zaal}, \mbox{16-10-10-10} and \mbox{16-16-10-10} ANN designs trained using {\sc pytorch}, respectively. Observe that the reduction on the hardware complexity of ANN designs under the parallel architecture is greater than that of the ANN designs under the time-multiplexed architectures due to a large number of multipliers and adders under the parallel architecture.

\begin{figure*}[t]
	\centerline{\includegraphics[width=18.5cm]{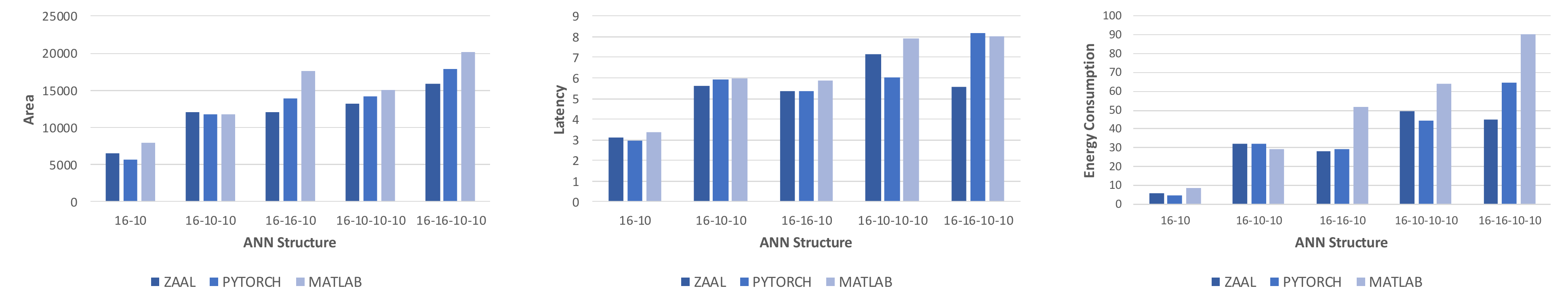}}
	\vspace*{-2mm}
	\caption{Multiplierless ANN designs under the parallel architecture using CAVM blocks after the post-training phase.}
	\label{fig:mul_cavm_parallel}
	\vspace*{-4mm}
\end{figure*}

\begin{figure*}[t]
	\centerline{\includegraphics[width=18.5cm]{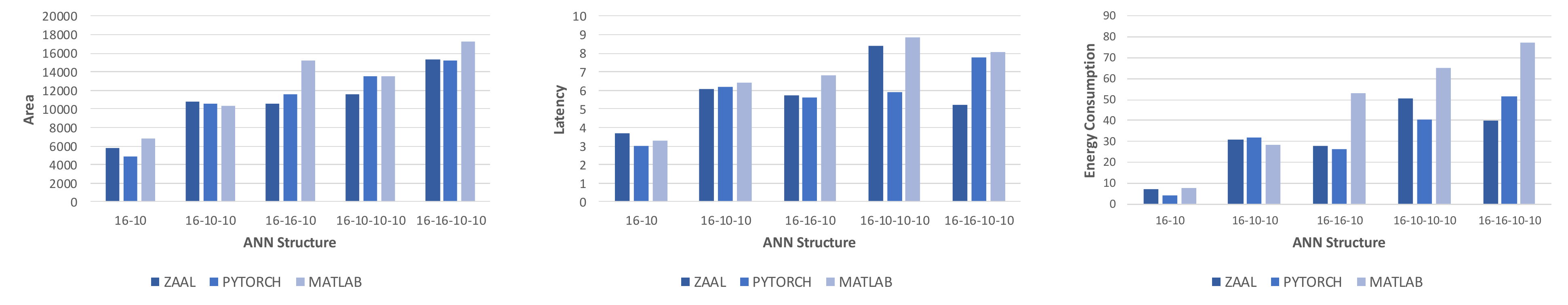}}
	\vspace*{-2mm}
	\caption{Multiplierless ANN designs under the parallel architecture using CMVM blocks after the post-training phase.}
	\label{fig:mul_cmvm_parallel}
	\vspace*{-4mm}
\end{figure*}

\begin{figure*}[t]
	\centerline{\includegraphics[width=18.5cm]{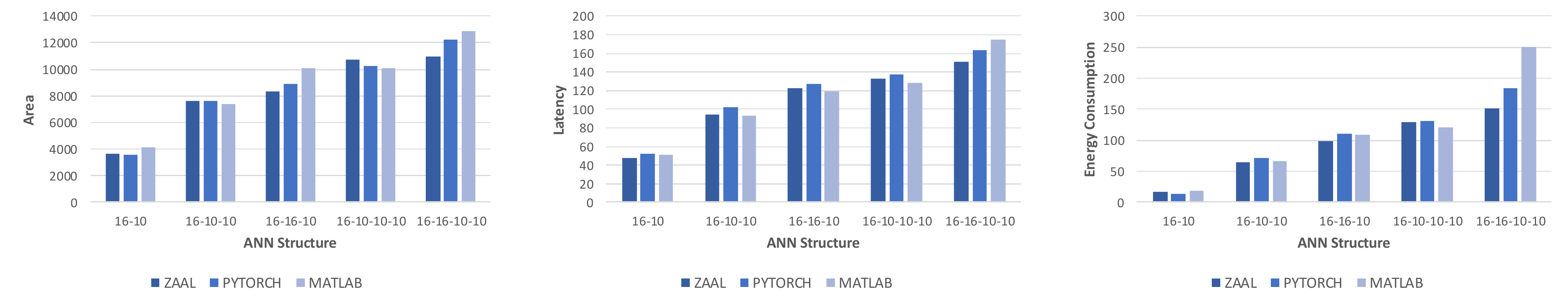}}
	\vspace*{-2mm}
	\caption{Multiplierless ANN designs under the {\sc smac\_neuron} architecture using MCM blocks after the post-training phase.}
	\label{fig:mul_smac_neuron}
	\vspace*{-6mm}
\end{figure*}

In order to compare the impact of the multiplierless design on the ANN hardware complexity, Figs.~\ref{fig:mul_cavm_parallel}-\ref{fig:mul_smac_neuron} present respectively gate-level results of multiplierless ANN designs under the parallel architecture using CAVM and CMVM blocks, and of multiplierless ANN designs under the {\sc smac\_neuron} architecture using MCM blocks after the post-training phase. 


Observe from Figs.~\ref{fig:tuning_parallel} and~\ref{fig:mul_cavm_parallel}-\ref{fig:mul_cmvm_parallel} that the multiplierless realization of CAVM and CMVM blocks reduces the area of ANN designs under the parallel architecture significantly. When compared to ANN designs where constant multiplications are described in a behavioral fashion after the post-training phase, the multiplierless design of ANNs using CAVM blocks leads to a maximum 11\% area reduction obtained on the \mbox{16-10-10} ANN design trained using {\sc pytorch}. Note that the multiplierless realization of ANNs using CMVM blocks leads to a larger area reduction than those using CAVM blocks. This is due to the fact that the sharing of subexpressions is increased in the CMVM block, reducing the total number of adders and subtractors. When compared to ANN designs where constant multiplications are described in a behavioral fashion after the post-training phase, the multiplierless design of ANNs using CMVM blocks leads to a maximum 28\% area reduction obtained on the \mbox{16-16-10} ANN design trained using {\sc matlab}. Moreover, the multiplierless design of ANNs using CMVM blocks reduces the energy consumption on average. However, in the multiplierless ANN designs, the latency is increased due to a large number of adders and subtractors in series in CAVM and CMVM blocks. On the other hand, similar results are also obtained on the multiplierless ANN design under the {\sc smac\_neuron} architecture as can be observed from Figs.~\ref{fig:tuning_smac_neuron} and~\ref{fig:mul_smac_neuron}. When compared to the ANN design where constant multiplications are described in a behavioral fashion after the post-training phase, the multiplierless design of ANNs under the {\sc smac\_neuron} architecture achieves a maximum 20\% area reduction obtained on the \mbox{16-10-10-10} ANN design trained using {\sc matlab}. 

We note that proposed post-training and multiplierless design techniques can reduce the ANN hardware complexity independently of the training algorithm, although different training methods lead to ANN designs with different hardware complexity and accuracy.

\section{Conclusions}
\label{sec:conclusions}

This article presented alternative implementations of feedforward ANNs under the parallel and time-multiplexed architectures. For each architecture, it also introduced a hardware-aware post-training phase where weight and bias values of ANNs are tuned in order to reduce the hardware complexity taking into account the hardware accuracy. Moreover, it proposed design techniques to implement ANNs under the shift-adds architecture without using multipliers. Furthermore, it introduced a CAD tool that can automatically generate the hardware description of the ANN design under the given ANN specifications. Experimental results indicate that proposed post-training methods can reduce the ANN hardware complexity significantly and the multiplierless design of ANNs can reduce the area and energy consumption further, increasing the latency slightly.
\section*{Acknowledgment}

This work is supported by the TUBITAK-Career project \#119E507.

\bibliographystyle{IEEEtran}
\bibliography{tvlsi20}

\begin{thebibliography}{10}
\providecommand{\url}[1]{#1}
\csname url@samestyle\endcsname
\providecommand{\newblock}{\relax}
\providecommand{\bibinfo}[2]{#2}
\providecommand{\BIBentrySTDinterwordspacing}{\spaceskip=0pt\relax}
\providecommand{\BIBentryALTinterwordstretchfactor}{4}
\providecommand{\BIBentryALTinterwordspacing}{\spaceskip=\fontdimen2\font plus
\BIBentryALTinterwordstretchfactor\fontdimen3\font minus
  \fontdimen4\font\relax}
\providecommand{\BIBforeignlanguage}[2]{{%
\expandafter\ifx\csname l@#1\endcsname\relax
\typeout{** WARNING: IEEEtran.bst: No hyphenation pattern has been}%
\typeout{** loaded for the language `#1'. Using the pattern for}%
\typeout{** the default language instead.}%
\else
\language=\csname l@#1\endcsname
\fi
#2}}
\providecommand{\BIBdecl}{\relax}
\BIBdecl

\bibitem{medical_image}
Q.~{Li}, W.~{Cai}, X.~{Wang}, Y.~{Zhou}, D.~D. {Feng}, and M.~{Chen}, ``Medical
  image classification with convolutional neural network,'' in
  \emph{International Conference on Control Automation Robotics Vision}, 2014,
  pp. 844--848.

\bibitem{face_detection}
H.~Li, Z.~Lin, X.~Shen, J.~Brandt, and G.~Hua, ``A convolutional neural network
  cascade for face detection,'' in \emph{IEEE Conference on Computer Vision and
  Pattern Recognition}, 2015, pp. 5325--5334.

\bibitem{alexnet}
A.~Krizhevsky, I.~Sutskever, and G.~E. Hinton, ``Imagenet classification with
  deep convolutional neural networks,'' in \emph{International Conference on
  Neural Information Processing Systems}, 2012, pp. 1106--1114.

\bibitem{misra10}
J.~Misra and I.~Saha, ``Artificial neural networks in hardware: A survey of two
  decades of progress,'' \emph{Neurocomputing}, vol.~74, no. 1-3, pp. 239--255,
  2010.

\bibitem{hecht-nielsen90}
R.~Hecht-Nielsen, \emph{Neurocomputing}.\hskip 1em plus 0.5em minus 0.4em\relax
  Addison-Wesley, 1990.

\bibitem{haykin99}
S.~Haykin, \emph{Neural Networks: A Comprehensive Foundation}.\hskip 1em plus
  0.5em minus 0.4em\relax Prentice-Hall, 1999.

\bibitem{courbariaux15}
M.~Courbariaux, Y.~Bengio, and J.-P. David, ``Binaryconnect: Training deep
  neural networks with binary weights during propagations,'' in
  \emph{International Conference on Neural Information Processing Systems},
  2015, pp. 3123--3131.

\bibitem{courbariaux16}
M.~Courbariaux, I.~Hubara, D.~Soudry, R.~El-Yaniv, and Y.~Bengio, ``Binarized
  neural networks: Training deep neural networks with weights and activations
  constrained to +1 or -1,'' \emph{arXiv e-prints}, 2016, arXiv:1602.02830.

\bibitem{ding18}
R.~Ding, Z.~Liu, R.~D. Blanton, and D.~Marculescu, ``Quantized deep neural
  networks for energy efficient hardware-based inference,'' in \emph{Asia and
  South Pacific Design Automation Conference}, 2018, pp. 1--8.

\bibitem{hashemi17}
H.~{Tann}, S.~{Hashemi}, R.~I. {Bahar}, and S.~{Reda}, ``Hardware-software
  codesign of accurate, multiplier-free deep neural networks,'' in \emph{Design
  Automation Conference}, 2017, pp. 28:1--28:6.

\bibitem{hybrid-ann}
H.~Park and T.~Kim, ``Structure optimizations of neuromorphic computing
  architectures for deep neural network,'' in \emph{Design, Automation and Test
  in Europe Conference and Exhibition}, 2018, pp. 183--188.

\bibitem{sarwar16}
S.~S. Sarwar, S.~Venkataramani, A.~Raghunathan, and K.~Roy, ``Multiplier-less
  artificial neurons exploiting error resiliency for energy-efficient neural
  computing,'' in \emph{Design, Automation and Test in Europe Conference and
  Exhibition}, 2016, pp. 145--150.

\bibitem{szabo00}
T.~Szabo, L.~Antoni, G.~Horvath, and B.~Feher, ``{A full-parallel digital
  implementation for pre-trained NNs},'' in \emph{International Joint
  Conference on Neural Networks}, 2000, pp. 49--54.

\bibitem{aksoy_iscas20}
L.~Aksoy, S.~Parvin, M.~E. Nojehdeh, and M.~Altun, ``Efficient time-multiplexed
  realization of feedforward artificial neural networks,'' in
  \emph{International Symposium on Circuits and Systems}, 2020, accepted for
  publication.

\bibitem{parhi99}
K.~Parhi, \emph{VLSI Digital Signal Processing Systems: Design and
  Implementation}.\hskip 1em plus 0.5em minus 0.4em\relax John Wiley \& Sons,
  1999.

\bibitem{horowitz14}
M.~Horowitz, ``Computing's energy problem (and what we can do about it),'' in
  \emph{IEEE International Solid-State Circuits Conference (ISSCC)}, 2014.

\bibitem{aksoy10}
L.~Aksoy, E.~O. Gunes, and P.~Flores, ``{Search algorithms for the multiple
  constant multiplications problem: Exact and approximate},''
  \emph{Microprocessors and Microsystems, Embedded Hardware Design}, vol.~34,
  no.~5, pp. 151--162, 2010.

\bibitem{aksoy12}
L.~Aksoy, E.~Costa, P.~Flores, and J.~Monteiro, ``Multiplierless design of
  linear {DSP} transforms,'' in \emph{VLSI-SoC: Advanced Research for Systems
  on Chip}, S.~Mir, C.-Y. Tsui, R.~Reis, and O.~C.~S. Choy, Eds.\hskip 1em plus
  0.5em minus 0.4em\relax Springer Berlin Heidelberg, 2012, pp. 73--93.

\bibitem{aksoy_iscas14}
L.~Aksoy, P.~Flores, and J.~Monteiro, ``{ECHO:} {A} novel method for the
  multiplierless design of constant array vector multiplication,'' in
  \emph{International Symposium on Circuits and Systems}, 2014, pp. 1456--1459.

\bibitem{pytorch}
A.~Paszke, S.~Gross, S.~Chintala, G.~Chanan, E.~Yang, Z.~DeVito, Z.~Lin,
  A.~Desmaison, L.~Antiga, and A.~Lerer, ``Automatic differentiation in
  pytorch,'' in \emph{Conference on Neural Information Processing Systems,
  Autodiff Workshop}, 2017.

\bibitem{matlab-ann}
\BIBentryALTinterwordspacing
{The MathWorks Inc.}, \emph{Deep Learning Toolbox}, Natick, Massachusetts,
  United States, 2020. [Online]. Available:
  \url{https://www.mathworks.com/help/deeplearning/}
\BIBentrySTDinterwordspacing

\bibitem{oscar07}
O.~Gustafsson, ``Lower bounds for constant multiplication problems,''
  \emph{IEEE Transactions on Circuits and Systems II: Express Briefs}, vol.~54,
  no.~11, pp. 974--978, 2007.

\bibitem{ercegovac}
M.~Ercegovac and T.~Lang, \emph{Digital Arithmetic}.\hskip 1em plus 0.5em minus
  0.4em\relax Morgan Kaufmann, 2003.

\bibitem{boullis05}
N.~Boullis and A.~Tisserand, ``Some optimizations of hardware multiplication by
  constant matrices,'' \emph{IEEE Transactions on Computers}, vol.~54, no.~10,
  pp. 1271--1282, 2005.

\bibitem{gustafsson07}
O.~Gustafsson, ``A difference based adder graph heuristic for multiple constant
  multiplication problems,'' in \emph{International Symposium on Circuits and
  Systems}, 2007, pp. 1097--1100.

\bibitem{voronenko07}
M.~P. Y.~Voronenko, ``Multiplierless multiple constant multiplication,''
  \emph{ACM Transactions on Algorithms}, vol.~3, no.~2, 2007.

\bibitem{kang01}
H.~J. Kang and I.~C. Park, ``{FIR} filter synthesis algorithms for minimizing
  the delay and the number of adders,'' \emph{IEEE Transactions on Circuits and
  Systems II: Analog and Digital Signal Processing}, vol.~48, no.~8, pp.
  770--777, 2001.

\bibitem{demirsoy02}
S.~S. Demirsoy, A.~G. Dempster, and I.~Kale, ``Power analysis of multiplier
  blocks,'' in \emph{International Symposium on Circuits and Systems}, 2002,
  pp. 297--300.

\bibitem{aksoy_dsd10}
L.~Aksoy, E.~Costa, P.~Flores, and J.~Monteiro, ``Optimization of area and
  delay at gate-level in multiple constant multiplications,'' in
  \emph{Euromicro Conference on Digital System Design, Architectures, Methods
  and Tools}, 2010, pp. 3--10.

\bibitem{kumm17}
M.~Kumm, M.~Hardieck, and P.~Zipf, ``Optimization of constant matrix
  multiplication with low power and high throughput,'' \emph{{IEEE}
  Transactions on Computers}, vol.~66, no.~12, pp. 2072--2080, 2017.

\bibitem{demirsoy07}
S.~Demirsoy, I.~Kale, and A.~Dempster, ``Reconfigurable multiplier constant
  blocks: Structures, algorithm and applications,'' \emph{Springer Circuits,
  Systems and Signal Processing}, vol.~26, no.~6, pp. 793--827, 2007.

\bibitem{aksoy14}
L.~Aksoy, P.~Flores, and J.~Monteiro, ``{Multiplierless design of folded DSP
  blocks},'' \emph{ACM Transactions on Design Automation of Electronic
  Systems}, vol.~20, no.~1, pp. 14:1--14:24, 2014.

\bibitem{moller16}
K.~Möller, M.~Kumm, M.~Kleinlein, and P.~Zipf, ``{Reconfigurable constant
  multiplication for FPGAs},'' \emph{IEEE Transactions on Computer-Aided Design
  of Integrated Circuits and Systems}, vol.~36, no.~6, pp. 927--937, 2016.

\bibitem{modified-mac}
Y.~Seo and D.~Kim, ``A new vlsi architecture of parallel
  multiplier–accumulator based on radix-2 modified booth algorithm,''
  \emph{IEEE Transactions on Very Large Scale Integration (VLSI) Systems},
  vol.~18, no.~2, pp. 201--208, 2010.

\bibitem{fpga-mac}
N.~Nedjah, R.~M. da~Silva, L.~M. Mourelle, and M.~V.~C. da~Silva, ``{Dynamic
  MAC-based architecture of artificial neural networks suitable for hardware
  implementation on FPGAs},'' \emph{Neurocomputing}, vol.~72, no.~10, pp. 2171
  -- 2179, 2009.

\bibitem{adam14}
D.~P. {Kingma} and J.~{Ba}, ``Adam: A method for stochastic optimization,''
  \emph{arXiv e-prints}, 2014, arXiv:1412.6980.

\bibitem{xavier10}
X.~Glorot and Y.~Bengio, ``Understanding the difficulty of training deep
  feedforward neural networks,'' in \emph{International Conference on
  Artificial Intelligence and Statistics}, 2010, pp. 249--256.

\bibitem{he15}
K.~{He}, X.~{Zhang}, S.~{Ren}, and J.~{Sun}, ``Delving deep into rectifiers:
  Surpassing human-level performance on imagenet classification,'' \emph{arXiv
  e-prints}, 2015, arXiv:1502.01852.

\bibitem{nwankpa18}
C.~Nwankpa, W.~Ijomah, A.~Gachagan, and S.~Marshall, ``Activation functions:
  Comparison of trends in practice and research for deep learning,''
  \emph{arXiv e-prints}, 2018, arXiv:1811.03378.

\bibitem{alimoglu97}
F.~Alimoglu and E.~Alpaydin, ``Combining multiple representations and
  classifiers for pen-based handwritten digit recognition,'' in
  \emph{International Conference on Document Analysis and Recognition}, 1997,
  pp. 637--640.

\end{thebibliography}

\end{document}